\documentclass[prd,twocolumn]{revtex4}%
\usepackage{amsfonts}
\usepackage{amsmath}
\usepackage{amssymb}
\usepackage{graphicx}%
\begin{document}
\title{Anti-Unruh effect in the thermal background }
\author{Yongjie Pan}
\author{Baocheng Zhang}
\email{zhangbc.zhang@yahoo.com}
\affiliation{School of Mathematics and Physics, China University of Geosciences, Wuhan
430074, China}

\begin{abstract}
We study the influence of the thermal background on the existence of the
anti-Unruh effect. For the massless scalar field, we present that the
anti-Unruh effect can appear when the detector is accelerated in the thermal
field, which is shown to be absent when accelerating in the Minkowski vacuum.
For the massive scalar field, it is found that the anti-Unruh effect which
exists in the case for accelerated detectors in the vacuum can disappear when
the background temperature increases. We use the many-body entangled quantum
state to study the situation of the massive scalar field. When the many-body
state is accelerated, its entanglement is decreasing with the increase of the
background temperature and the phenomenon of the sudden death for entanglement
occurs. This provides a valuable indication that the anti-Unruh effect can
exist below some specific background temperature, dependent on the actual
experimental environment.

\end{abstract}
\maketitle

\section{Introduction}

The Unruh effect \cite{wgu76} was discovered in 1976, which states that an
observer with uniform acceleration would feel a thermal bath of particles in
the Minkowski vacuum of a free quantum field. Since then, the effect had been
digested and extended to many different situations (see the review
\cite{chm08} and references therein). Although many proposals based mainly on
the Unruh-DeWitt (UDW) detector \cite{bsd79} had been put forward
\cite{bl83,uw84,vm01,sy03,ssh06,oyz16,ccm16,clv17,ltc17,lck19,hfc19,lbv19,blp20}
in past years, the observation of Unruh effect has not been realized up to
now, because of the pretty low Unruh temperature which requires the
acceleration to be about $10^{20}m/s^{2}$ in order to realize a photon bath at
$1K$. As is well known, an elemental obstacle to test the Unruh effect
experimentally is the thermal noise from the environment. The recent found
anti-Unruh effect \cite{bmm16} provides a helpful way to test the effect
experimentally, since it usually leads to the different behaviors from that by
the thermal noise. This phenomena of anti-Unruh effect follows from the
observation \cite{bmm16} that an increase in detector acceleration will
correspond to a decrease in the temperature of the detected radiation, in
direct contrast to both the Unruh effect and one's expectations. In
particular, the anti-Unruh effect has been shown to represent a general
stationary mechanism that can exist under a stationary state satisfying
Kubo-Martin-Schwinger (KMS) condition \cite{rk57,ms59,fjl16} and is
independent on any kind of boundary conditions \cite{gmr16}. Thus, like the
Unruh effect, the anti-Unruh effect constitutes another new phenomenon for the
accelerated observers.

Since the experiments are always made in the range of finite length and time,
it must distinguish the two situations of Unruh and anti-Unruh effects
carefully. An interesting way for this is to see the change of quantum
entanglement by acceleration. According to the previous results
\cite{fm05,amt06,ml09,mgl10,wj11,ses12,bfl12,ro15,dss15}, the quantum
entanglement would be degraded by the Unruh effect, which is similar to the
results caused by the thermal background. However, a recent calculation showed
that the anti-Unruh effect can lead to the increase for the quantum
entanglement for the bipartite \cite{lzy18} and many-body quantum states
\cite{pz20}.

The previous studies was mostly made for the change of quantum states and
entanglement when the physical systems were accelerated in the Minkowski
vacuum. As the thermal noise cannot be eliminated completely, it is
significant to investigate the influence of acceleration on the quantum states
or entanglement in the thermal background \cite{cm95,sk14,mal20,bm21,bbm21}.
Since it is difficult to distinguish the physical results by the Unruh effect
from that by the thermal background \cite{kp15,cdm19,lal20}, we will
investigate the influence of anti-Unruh effect on many-body entanglement and
study whether it could be differentiated from thermal background in this
paper. We will consider the spin squeezed states \cite{ku93,mws11} as done
before in Ref. \cite{pz20} but accelerated in the thermal background instead
of the vacuum background. This is necessary because for accelerated observers
in thermal field the Green functions which are essential in calculating the
transition probability are not time translational invariant and the
accelerated observers see the thermal background not in thermal equilibrium
\cite{cdm19,lal20}. Thus, whether the earlier investigation of the anti-Unruh
effect for accelerated many-body quantum systems in vacuum is still hold in
the thermal background is unclear both in mathematical and physical
perspective. This constitutes the main motivation for the present work. In
this paper, we not only investigate the influence of background thermal field
on the existence of anti-Unruh effect, but also we will study the influence on
the accelerated many-body quantum systems. We will focus on twin-Fock (TF)
states \cite{bh93} which can be seen as a kind of limit for spin squeezed
states and had been realized in a recent experiment with more than $10^{4}$
atoms \cite{lzy17}.

This paper is organized as follows. In the second section we review the
two-level Unruh-DeWitt (UDW) detector model, and investigate whether the
anti-Unruh effect appears for the environment of the vacuum or the thermal
field with the condition of the massless scalar field. Then, the model for the
atom accelerated in the thermal field is given in the third section. This is
followed in the fourth section by the discussions on the influence of
acceleration on entanglement for TF states in the thermal background, where
the spin squeezing parameter is used to measure the change of entanglement.
Then, the phase sensitivity is analyzed for the accelerated TF states in the
thermal background with the actual experimental conditions in Sec. IV.
Finally, we give a conclusion in Sec. V. In this paper, we use units with
$c=\hbar=k_{B}=1$, except the part of analyzing the phase sensitivity in Sec. IV.

\section{Thermal Field}

In this section we will investigate the influence of thermal field on
existence of anti-Unruh effect qualitatively. Start from the model of the UDW
detector. It is usually considered as a pointlike two-level quantum system or
atom (as required in this paper) and consists of two quantum states, i.e., the
ground $\left\vert g\right\rangle $ and excited $\left\vert e\right\rangle $
states, which are separated by an energy gap $\Omega$ while experiencing
accelerated motion in a vacuum field. But for the accelerated atom, the Unruh
effect will influence the state of the atom. This could be described according
to the following interaction Hamiltonian in a ($1+1$)-dimension model,
$H_{I}=\lambda\chi\left(  \tau/\sigma\right)  \mu\left(  \tau\right)
\phi\left(  x\left(  \tau\right)  \right)  $, where $\phi$ is a minimally
coupled scalar field related to the thermal background in Minkowski spacetime
and interacts with the accelerated atom, $\lambda$ is the coupling strength,
$\tau$ is the atom's proper time along its trajectory $x\left(  \tau\right)
$, $\mu\left(  \tau\right)  $ is the atom's monopole momentum, and
$\chi\left(  \tau/\sigma\right)  $ is a switching function that is used to
control the interaction timescale $\sigma$. Thus, the total excitation
probability for an accelerated detector initially in the ground state and
evolving in the thermal background characterized by a temperature $\beta
^{-1}=T$ can be expressed as \cite{cm95,bm21}%
\begin{equation}
P_{+}=C\int d\tau\int d\tau^{\prime}\chi\left(  \frac{\tau}{\sigma}\right)
\chi\left(  \frac{\tau^{\prime}}{\sigma}\right)  e^{-i\Omega\left(  \tau
-\tau^{\prime}\right)  }G_{\beta}^{+}\left(  \tau,\tau^{\prime}\right)
\label{pgt}%
\end{equation}
which is calculated only at the leading order with $C=\lambda^{2}\left\vert
\left\langle e\right\vert \mu\left(  0\right)  \left\vert g\right\rangle
\right\vert ^{2}$. $G_{\beta}^{+}\left(  \tau,\tau^{\prime}\right)  \equiv
G_{\beta}^{+}\left(  x\left(  \tau\right)  ,x\left(  \tau^{\prime}\right)
\right)  =\left\langle \phi\left(  x\left(  \tau\right)  \right)  \phi\left(
x\left(  \tau^{\prime}\right)  \right)  \right\rangle _{\beta}$ is the thermal
Green function by taking the Gibbs ensemble average. When $T=0$, $G_{\beta
}^{+}\left(  \tau,\tau^{\prime}\right)  =G_{0}^{+}\left(  \tau,\tau^{\prime
}\right)  $ which is the vacuum Green function for accelerating atom in the vacuum.

At first, we make a little analysis about the influence of the time
transitional invariance on the existence of anti-Unruh effect. As known in
Ref. \cite{gmr16,cdm19} that $G_{0}^{+}\left(  \tau,\tau^{\prime}\right)
=G_{0}^{+}\left(  \tau-\tau^{\prime}\right)  $ which indicates that the vacuum
Green function is time transitional invariant, but the entire integrand
$F_{0}\left(  \tau,\tau^{\prime}\right)  =\chi\left(  \frac{\tau}{\sigma
}\right)  \chi\left(  \frac{\tau^{\prime}}{\sigma}\right)  e^{-i\Omega\left(
\tau-\tau^{\prime}\right)  }G_{0}^{+}\left(  x\left(  \tau\right)  ,x\left(
\tau^{\prime}\right)  \right)  $ is not invariant because the switching
function takes the Gauss form, $\chi\left(  \frac{\tau}{\sigma}\right)
=\exp[-\frac{\tau^{2}}{2\sigma^{2}}]$. We can choose the proper switching
function to preserve the time transitional invariance for the integrand
$F_{0}\left(  \tau,\tau^{\prime}\right)  $, i.e. $\chi\left(  \frac{\tau
}{\sigma}\right)  =1$. From the Fig. 1, it is seen that the anti-Unruh effect
exists for the two situations, which shows at least that the time transitional
invariance of the vacuum Green function is not conclusive element for the
existence of the anti-Unruh effect. Furthermore, it might imply that the
anti-Unruh effect could exist in the thermal background even though the
thermal Green function is not time transitional invariance. In this paper, we
will show this point by two different aspects. The one is to present that the
anti-Unruh effect could appear in thermal background but it does not exist in
the vacuum for the massless scalar field. Another one is to show that the
anti-Unruh effect exists in the vacuum for the massive scalar field but
whether it still exists in thermal background depends on the temperature of
the background field. The latter one will be studied in the next section with
the many-body quantum system, and the former one will be analyzed here qualitatively.

\begin{figure}[ptb]
\centering
\includegraphics[width=1\columnwidth]{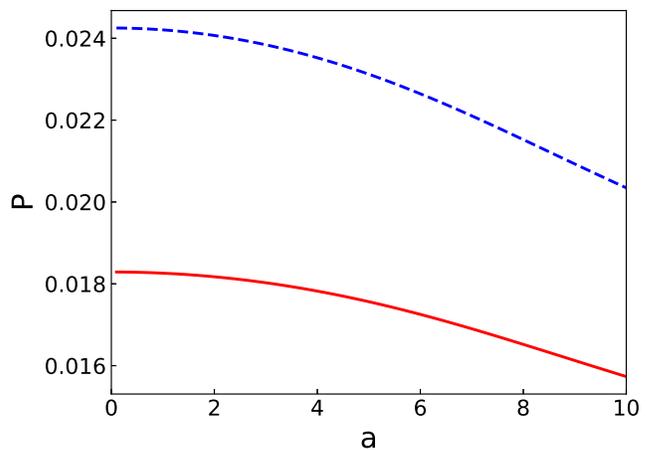} \caption{(Color online) The
transition probability as a function of the acceleration $a$. The model
parameters employed are $\lambda= 1, \sigma= 0.2, \Omega= 0.3$. The solid red
line denotes the case in which the switching function is Gaussian, and the
dashed blue line denotes the case in which the switching function is
rectangular due to the finite interaction time.}%
\label{Fig1}%
\end{figure}

When the accelerated atom moves along the trajectory,%
\begin{align}
t\left(  \tau\right)   &  =a^{-1}\sinh(a\tau),\nonumber\\
x\left(  \tau\right)   &  =a^{-1}\cosh(a\tau),\label{ast}%
\end{align}
with the proper acceleration $a$, ones can calculate the Green functions as
\cite{cm95,lal20}%
\begin{equation}
G_{0}^{+}\left(  \tau,0\right)  =-\frac{a^{2}}{16\pi^{2}}\frac{1}{\sinh
^{2}(a\tau/2)},\label{vgf}%
\end{equation}
for the vacuum, and
\begin{align}
G_{\beta}^{+}\left(  \tau,0\right)   &  =\frac{a}{16\pi\beta\sinh^{2}%
(a\tau/2)}[\coth(\frac{\pi}{a\beta}(e^{a\tau}-1))\nonumber\\
&  -\coth(\frac{\pi}{a\beta}(1-e^{-a\tau}))],\label{tgf}%
\end{align}
for the thermal field. Here we take $\tau^{\prime}=0$ for simplicity. Note
that the breakdown of the time transitional invariance in Eq. (\ref{tgf})
cannot be seen clearly. It does not matter for the analysis of the anti-Unruh
effect, as pointed out above. Moreover, the thermal Green function (\ref{tgf})
is calculated using massless scalar field, and will reduce to the vacuum
result (\ref{vgf}) as the background temperature approaches to zero or
$\beta\rightarrow\infty$.

According to Ref. \cite{gmr16}, the anti-Unruh effect exists under the weak
condition, $\frac{\partial P_{+}}{\partial a}<0$, base on such concept for
anti-Unruh effect: the transition probability of an accelerated detector can
actually decrease with acceleration. As analyzed above, the switching function
is not crucial for the existence of the anti-Unruh effect. Thus, we ignore the
switching function here for the further analysis. This means that $P_{+}\sim
C\int d\tau e^{-i\Omega\tau}G^{+}\left(  \tau,0\right)  \sim C\widetilde
{G}\left(  \Omega\right)  $ where $\widetilde{G}\left(  \Omega\right)  $ is
the Fourier transform of $G^{+}\left(  \tau,0\right)  $. Therefore, the weak
condition for the anti-Unruh effect becomes $\frac{\partial\widetilde
{G}\left(  \Omega\right)  }{\partial a}<0$.

For the acceleration in the vacuum, we have%
\begin{equation}
\widetilde{G}_{0}\left(  \Omega\right)  =\frac{a+\pi\Omega\coth\left(
\pi\Omega/a\right)  }{2\pi^{5/2}}.
\end{equation}
It is not hard to show that $\frac{\partial\widetilde{G}_{0}\left(
\Omega\right)  }{\partial a}>0$ for any $a$ or $\Omega$. This is just the
result in Ref. \cite{gmr16} that the anti-Unruh effect is absent for
accelerated detectors in the Minkowski vacuum of a massless scalar field.

For the acceleration in the thermal field, we can calculate the transition
probability as in the case for the acceleration in the vacuum. However, there
is no analytical expression for the Fourier transform of $G_{\beta}^{+}\left(
\tau,0\right)  $ in Eq. (\ref{tgf}). Therefore, the numerical result is given
in Fig. 2, in which it is seen that anti-Unruh effect appears for the small
acceleration. This shows that the introduction of thermal field would change
the influence of acceleration on the quantum transition process. In the next
section, we will discuss that the influence of thermal field on the change of
quantum states or entanglement led by acceleration with the many-body physical systems.

\begin{figure}[ptb]
\centering
\includegraphics[width=1\columnwidth]{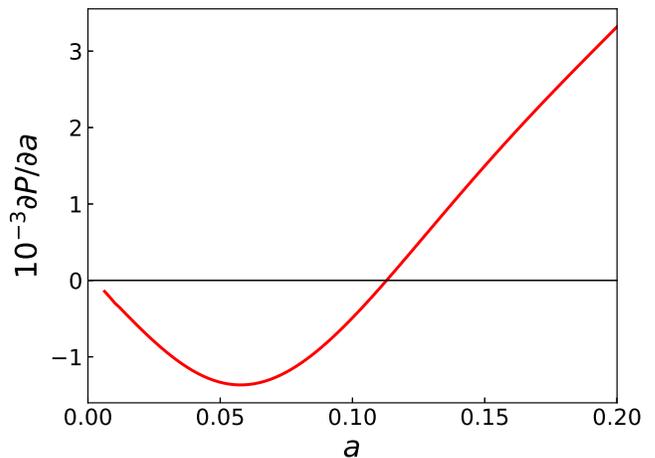} \caption{(Color online) The
derivative of the transition probability with respect to the acceleration as a
function of the acceleration $a$. The inverse of the background temperature is
taken as $\beta= 30$, and the other parameters are the same as in Fig. 1. }%
\label{Fig2}%
\end{figure}

\section{Acceleration in thermal field}

We still consider the UDW model for the single atom and the interaction
unitary operation $U=I-i\int d\tau H_{I}\left(  \tau\right)  +O\left(
\lambda^{2}\right)  $ is expanded to the first order. As the atoms are
accelerated in thermal background, the evolution of the accelerated atom can
be written with the density operators as
\begin{equation}
\rho_{f}=U^{\dagger}\rho_{i}U,
\end{equation}
where the initial density operator consists of the product form of the density
operator for the atom and the density operator for the thermal field,
$\rho_{th}=\sum_{n}p_{n}\left\vert n\right\rangle \left\langle n\right\vert $
with $p_{n}=\left(  1-e^{-\beta\omega}\right)  e^{-n\beta\omega}$. Thus,
within the first-order approximation and in the interaction picture, the
evolution of the atom could be described by%
\begin{align}
Tr_{th}\left[  U^{\dagger}\left(  \rho_{a1}\otimes\rho_{th}\right)  U\right]
&  =\sum_{n}p_{n}\left(  \left\vert g\right\rangle \left\langle g\right\vert
+n\left\vert u_{+}\right\vert ^{2}\left\vert e\right\rangle \left\langle
e\right\vert \right)  ,\nonumber\\
Tr_{th}\left[  U^{\dagger}\left(  \rho_{a2}\otimes\rho_{th}\right)  U\right]
&  =\sum_{n}p_{n}\left\vert g\right\rangle \left\langle e\right\vert
,\nonumber\\
Tr_{th}\left[  U^{\dagger}\left(  \rho_{a3}\otimes\rho_{th}\right)  U\right]
&  =\sum_{n}p_{n}\left\vert e\right\rangle \left\langle g\right\vert
,\nonumber\\
Tr_{th}\left[  U^{\dagger}\left(  \rho_{a4}\otimes\rho_{th}\right)  U\right]
&  =\sum_{n}p_{n}\left(  \left(  n+1\right)  \left\vert u_{-}\right\vert
^{2}\left\vert g\right\rangle \left\langle g\right\vert +\left\vert
e\right\rangle \left\langle e\right\vert \right)  , \label{foe}%
\end{align}
where $\lambda=1$ is taken, $Tr_{th}$ represents the calculation of tracing
out the field degrees of freedom, and the normalization factors are not
written but are considered in the following calculations for the plots. The
initial density operators for the atom are taken as $\rho_{a1}=$ $\left\vert
g\right\rangle \left\langle g\right\vert $, $\rho_{a2}=\left\vert
g\right\rangle \left\langle e\right\vert $, $\rho_{a3}=\left\vert
e\right\rangle \left\langle g\right\vert $, $\rho_{a4}=\left\vert
e\right\rangle \left\langle e\right\vert $. These expressions in Eq.
(\ref{foe}) provide the elementary forms of the evolution for the general
state of the atoms. Since an atom is only considered to have two levels of
energy, the change of the field should be limited within these basis of
\{$\left\vert n-1\right\rangle $, $\left\vert n\right\rangle $, $\left\vert
n+1\right\rangle $\}, dependent on the absorption or emission of the photon
from the field by the atom. $u_{\pm}=\sum_{k}\int_{-\infty}^{\infty}%
\chi\left(  \tau/\sigma\right)  \exp[\pm i\Omega\tau+i\omega_{k}t\left(
\tau\right)  -ikx\left(  \tau\right)  ]{d\tau}$ are related to the motion of
the atom. With the trajectory of the atom in Eq. (\ref{ast}), the transition
probability can be calculated according to
\begin{align}
P_{+}  &  =\left\langle n-1\right\vert \left\langle e\right\vert U^{\dagger
}\left(  \rho_{a1}\otimes\rho_{th}\right)  U\left\vert e\right\rangle
\left\vert n-1\right\rangle \nonumber\\
&  =\sum_{n}np_{n}\left\vert u_{+}\right\vert ^{2}.
\end{align}
Comparing it with Eq. (\ref{pgt}), it is not difficult to confirm their
consistency by taking the field $\phi\left(  x\left(  \tau\right)  \right)
=\sum_{k}\frac{e^{-ikx\left(  \tau\right)  }}{\sqrt{2\omega_{k}}}\left(
a_{k}e^{-i\omega_{k}t\left(  \tau\right)  }+a_{k}^{\dagger}e^{i\omega
_{k}t\left(  \tau\right)  }\right)  $ where $k$ denotes the mode of the
($1+1$)-dimension scalar field with (bosonic) annihilation (creation) operator
$a_{k}$ ($a_{k}^{\dag}$), $a_{k}\left\vert n\right\rangle =\sqrt{n}\left\vert
n-1\right\rangle $ and $a_{k}^{\dag}\left\vert n\right\rangle =\sqrt
{n+1}\left\vert n+1\right\rangle $. Note that $P_{+}$ represents the
transition probability from the ground state to the excitation state.
Similarly, the transition probability $P_{-}$ from the excitation state to the
ground state can be calculated as $P_{-}=\left\langle n+1\right\vert
\left\langle g\right\vert U^{\dagger}\left(  \rho_{a4}\otimes\rho_{th}\right)
U\left\vert g\right\rangle \left\vert n+1\right\rangle =\sum_{n}\left(
n+1\right)  p_{n}\left\vert u_{-}\right\vert ^{2}.$

It was pointed out that the change of the quantum state, i.e. the transition
probability, is dependent on the concrete parameters like the interaction time
scale $\sigma$ and the energy gap $\Omega$ \cite{bmm16}. The probability would
decrease as the acceleration or the Unruh temperature increases under some
conditions, which makes the atom \textquotedblleft feel\textquotedblright%
\ cooler instead of warm up expected by the Unruh effect. Although the
physically essential reasons remain to be explored for the difference between
the anti-Unruh and Unruh effects, some important elements, like the
interaction time, the detector's energy gap, the mass of the quantum field,
etc, could be selected out to distinguish them operationally. See Fig. 3 for
the presentation of this effect in the thermal background. Here the massive
field with e.g. $\omega=\sqrt{k^{2}+m^{2}}$ is considered as in Ref.
\cite{gmr16} so that the anti-Unruh effect will not be constrained by the
finite interaction time and its validity can be extended to situations where
the detector is switched on adiabatically over an infinite long time. Without
loss of generality, $m=1$ is used for the numerical calculations. From Fig. 3,
it is seen that the existence of the thermal background does not seem to
affect the trend of the change for the two cases of Unruh and anti-Unruh
effects, although the interaction between the atom and the thermal field is
more complicated than that between the atom and the vacuum. Actually, the
anti-Unruh effect will disappear with the increase of the background
temperature due to the disappearance of entanglement, which can be seen below
with the discussions for many-body entanglement.

\begin{figure}[ptb]
\centering
\includegraphics[width=1\columnwidth]{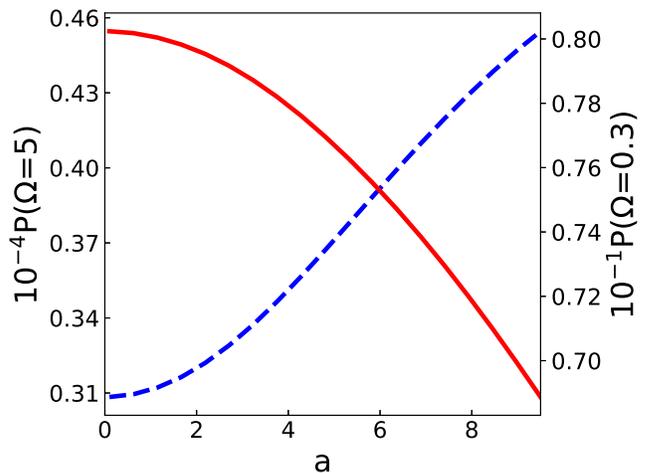} \caption{(Color online) The
transition probability as a function of the acceleration $a$ when the atom is
accelerated in the thermal field. The solid red line denotes the case that the
transition probability decreases with acceleration at $\Omega= 0.3$
(referenced to the right vertical axis), whereas the dashed blue line with
respect to the left vertical axis is for increasing transition probability
with acceleration at $\Omega= 5$. The inverse of the background temperature is
taken as $\beta= 1$, and the other parameters are the same as in Fig. 1. }%
\label{Fig3}%
\end{figure}

\section{Many-Body Entanglement}

We choose the TF states as the many-body entangled quantum states and discuss
the influence of acceleration on them in this section. TF states are one kind
of Dicke states \cite{rhd54}. For a collection of $N$ identical (pseudo-)
spin-1/2 particles, Dicke states can be expressed in Fock space as $\left\vert
\frac{N}{2}+m\right\rangle _{\uparrow}\left\vert \frac{N}{2}-m\right\rangle
_{\downarrow}$\ with ($\frac{N}{2}+m$) particles in spin-up and ($\frac{N}%
{2}-m$) particles in spin-down modes for $m=-\frac{N}{2},-\frac{N}{2}%
+1,\cdots,\frac{N}{2}$. In particular, $m=0$ represents just the TF state
where the number of the particles is the same for each one of the two spin
states. On the other hand, Dicke states can be described by the common
eigenstate $|j,m\rangle$ of the collective spin operators $J^{2}$ and $J_{z}$,
with respective eigenvalues $j(j+1)$ and $m$. For the system consisted of $N$
two-level atoms we will consider, the state $|j=\frac{N}{2},m\rangle$
indicates that ($j+m$) atoms are at the excited state $|e\rangle$, ($j-m$)
atoms are at the ground state $|g\rangle$. $J_{z}=\frac{1}{2}\left(
n_{e}-n_{g}\right)  $ represents the difference of the number of atoms between
excited ($n_{e}$) and ground ($n_{g}$) states, and $J^{2}=\frac{N}{2}\left(
\frac{N}{2}+1\right)  $ is related to the total number of atoms. For the TF
state, $m=0$ and $\left\langle J_{z}\right\rangle =0$.

The acceleration of TF states were investigated based on such consideration
that the single accelerated atom was calculated according to the UDW model and
all atoms were accelerated with the same way without any other interaction
among atoms except their initial entanglement. Meanwhile, the distance between
the atoms is much less than the relevant wavelengths of the field, which
assures that all atoms see the same field. When all atoms in the TF state are
accelerated, the state of every atom is changed. Because the excitation
probability is not equal to the deexcitation probability due to the
acceleration, the resulted atom's number at the excited state is not equal to
that at the ground state. Thus, the final state would deviate from the TF
state. In order to quantify the entanglement of the state, we choose the spin
squeezing parameter%
\begin{equation}
\xi_{E}^{2}=\frac{\left(  N-1\right)  \left(  \Delta J_{z}\right)
^{2}+\left\langle J_{z}^{2}\right\rangle }{\left\langle J^{2}\right\rangle
-N/2}. \label{sse}%
\end{equation}
It derives from the relation between spin squeezing and entanglement as shown
in \cite{sdp01,tkb09}, in which the inequality $\left(  N-1\right)  \left(
\Delta J_{\overrightarrow{n}}\right)  ^{2}+\left\langle J_{\overrightarrow{n}%
}^{2}\right\rangle \geqslant\left\langle J^{2}\right\rangle -N/2$ holds for
any separable states, and the violation of this inequality indicates
entanglement. Here the mean-spin direction $\overrightarrow{n}$ was taken
along the $z$ direction. If $\xi_{E}^{2}<1$, the state is spin squeezed and
entangled. In particular, the smaller the value of $\xi_{E}^{2}$, the more the
entanglement will be.

At first, we show the change of state for two accelerated atoms. It is noticed
that the bipartite quantum state for the maximal entangled atoms, $\left\vert
\psi_{i}\right\rangle =\frac{1}{\sqrt{2}}\left(  \left\vert g\right\rangle
_{A}\left\vert e\right\rangle _{B}+\left\vert e\right\rangle _{A}\left\vert
g\right\rangle _{B}\right)  $, can be regarded as the simplest TF state, i.e.
$\left\vert \psi_{i}\right\rangle =\left\vert 1,0\right\rangle $ using the
description in the above section. Here the subscripts $A$ and $B$ in the state
$\left\vert \psi_{i}\right\rangle $ represents the locations related to the
atoms $A$ and $B$. The initial density operator is assumed to take the form
\begin{equation}
\rho_{i}=\left\vert \psi_{i}\right\rangle \left\langle \psi_{i}\right\vert
\otimes\rho_{th}.
\end{equation}
For the case we consider, each atom is independently accelerating in the
thermal field and has the same coupling with the field in its respective
(spatial) place by the same process presented in Eq. (\ref{foe}). When the two
atoms are accelerated simultaneously in the thermal field, the state becomes
\begin{align}
&  \rho_{f}=Tr_{th}\left[  U_{A}^{\dagger}U_{B}^{\dagger}\rho_{i}U_{B}%
U_{A}\right] \nonumber\\
&  =\sum_{n_{A},n_{B}}\frac{p_{n_{A}}p_{n_{B}}}{D}[\frac{1}{2}\left\vert
u_{+}\right\vert ^{2}\left(  n_{A}+n_{B}\right)  \left\vert ee\right\rangle
\left\langle ee\right\vert \nonumber\\
&  +\frac{1}{2}\left(  1+n_{A}\left(  n_{B}+1\right)  \left\vert
u_{+}\right\vert ^{2}\left\vert u_{-}\right\vert ^{2}\right)  \left(
\left\vert ge\right\rangle +\left\vert eg\right\rangle \right)  \left(
\left\langle ge\right\vert +\left\langle eg\right\vert \right) \nonumber\\
&  +\frac{1}{2}\left\vert u_{-}\right\vert ^{2}\left(  n_{A}+n_{B}+2\right)
\left\vert gg\right\rangle \left\langle gg\right\vert ]
\end{align}
where $\left\vert ge\right\rangle \equiv\left\vert g\right\rangle
_{A}\left\vert e\right\rangle _{B}$ for simplicity, and $D=1+\frac{1}%
{2}\left\vert u_{+}\right\vert ^{2}\left(  n_{A}+n_{B}\right)  +\frac{1}%
{2}\left\vert u_{-}\right\vert ^{2}\left(  n_{A}+n_{B}+2\right)  +n_{A}\left(
n_{B}+1\right)  \left\vert u_{+}\right\vert ^{2}\left\vert u_{-}\right\vert
^{2}$ is the normalization factor. Furthermore, we write the final state
according to the representation for Dicke states,%
\begin{align}
\rho_{f}=  &  \sum_{n_{A},n_{B}}\frac{p_{n_{A}}p_{n_{B}}}{D}[\frac{1}%
{2}\left\vert u_{+}\right\vert ^{2}\left(  n_{A}+n_{B}\right)  \left\vert
1,1\right\rangle \left\langle 1,1\right\vert \nonumber\\
&  +\left(  1+n_{A}\left(  n_{B}+1\right)  \left\vert u_{+}\right\vert
^{2}\left\vert u_{-}\right\vert ^{2}\right)  \left\vert 1,0\right\rangle
\left\langle 1,0\right\vert \nonumber\\
&  +\frac{1}{2}\left\vert u_{-}\right\vert ^{2}\left(  n_{A}+n_{B}+2\right)
\left\vert 1,-1\right\rangle \left\langle 1,-1\right\vert .
\end{align}
Then, the difference of the number of atoms between excited and ground states
is obtained as%
\begin{align}
\left\langle J_{z}\right\rangle  &  =Tr\left(  \rho_{f}J_{z}\right)
\nonumber\\
&  =\sum_{n_{A},n_{B}}\frac{p_{n_{A}}p_{n_{B}}}{2D}[\left\vert u_{+}%
\right\vert ^{2}\left(  n_{A}+n_{B}\right) \nonumber\\
&  -\left\vert u_{-}\right\vert ^{2}\left(  n_{A}+n_{B}+2\right)  ],
\end{align}
where $Tr$ represents the trace of a matrix. The result means that the atom's
number at the excited state is not equal to that at the ground state,
different from the requirement of TF state, unless the probability of
transition from the ground state to the excited state equates the probability
for the inverse transition.

We extend the discussion for two atoms to the case of $N$ atoms with the
initial TF state $|j,0\rangle$. When all atoms are accelerated simultaneously
under thermal background, the TF state becomes%
\begin{equation}
\rho_{TF}=\sum_{m=-N/2}^{N/2}A_{m}^{2}|j,m\rangle\left\langle j,m\right\vert ,
\label{dsm}%
\end{equation}
up to the normalization factor which is included in our numerical calculation.
$A_{m}^{2}=[\sum_{k=0}^{N/2-\left\vert m\right\vert }C_{N/2}^{k}%
C_{N/2}^{k+\left\vert m\right\vert }\left(  P_{+}P_{-}\right)  ^{k}%
(\theta\left(  m\right)  \left(  P_{+}\right)  ^{m}+\theta\left(  -m\right)
\left(  P_{-}\right)  ^{\left\vert m\right\vert })]$ in which the function
$\theta\left(  x\right)  =1$ when $x>0$ and $\theta\left(  x\right)  =0$
otherwise, and $C_{n}^{r}=\frac{n!}{r!(n-r)!}$ denotes the combinatorial
factor of choosing $r$ out of $n$. The parameter $A_{0}^{2}$ represents the
probability of remaining the original form of the TF state, which includes
those cases that if $l$ $\left(  0\leqslant l\leqslant\frac{N}{2}\right)  $
atoms are changed from the ground states to the excited states, there must be
other $l$ atoms which are changed from the excited states to the ground states
simultaneously. The reason that the second term appears is due to the
inequality of the transition probabilities $P_{+}$ and $P_{-}$. The parameter
$A_{m}^{2}$ can be worked out by choosing the terms that in every term either
there are $m$ more excited states than ground states (that is the case for
$m>0$) or there are $m$ more ground states than excited states (that is the
case for $m<0$). The crossed terms like $|j,m\rangle\left\langle j,m^{\prime
}\right\vert $ have been reduced when tracing out the field degrees of freedom.

After acceleration, the TF state becomes $\rho_{TF}$ described in Eq.
(\ref{dsm}). With this, we can calculate $\left\langle J_{z}\right\rangle
=Tr\left(  \rho_{TF}J_{z}\right)  =\sum_{m=-N/2}^{N/2}mA_{m}^{2}$, and
$\left\langle J_{z}^{2}\right\rangle =Tr\left(  \rho_{TF}J_{z}^{2}\right)
=\sum_{m=-N/2}^{N/2}m^{2}A_{m}^{2}$. Thus, according to $\left(  \Delta
J_{z}\right)  ^{2}=\left\langle J_{z}^{2}\right\rangle -\left\langle
J_{z}\right\rangle ^{2}$, ones can calculate $\xi_{E}^{2}$ by substituting
these results into Eq. (\ref{sse}), which is presented in Fig. 4 for different
background temperatures. It is seen that the spin squeezing parameters for
three different background temperatures are decreasing or entanglement is
increasing when the acceleration is increased, which are the indication for
the anti-Unruh effect. It is noted that the entanglement at $a=0$ for the
accelerated state (\ref{dsm}) is less than that for the initial maximal
entangled state due to the presence of switching function.

Moreover, we calculate the change of $\xi_{E}^{2}$ with regard to the
background temperatures, which is presented in Fig. 5. It shows that the
entanglement is decreasing when the background temperature increases for a
fixed acceleration. When the background temperature increases to some specific
value, entanglement would decrease to zero, which is similar to the sudden
death of entanglement due to the increase of the temperature \cite{ro15}. So,
the observation of the anti-Unruh effect requires a lower background
temperature, since a high background temperature would change the influence of
acceleration on quantum many-body systems. Even a higher temperature would
eliminate the evidence for the influence of acceleration as entanglement
becomes zero.

\begin{figure}[ptb]
\centering
\includegraphics[width=1\columnwidth]{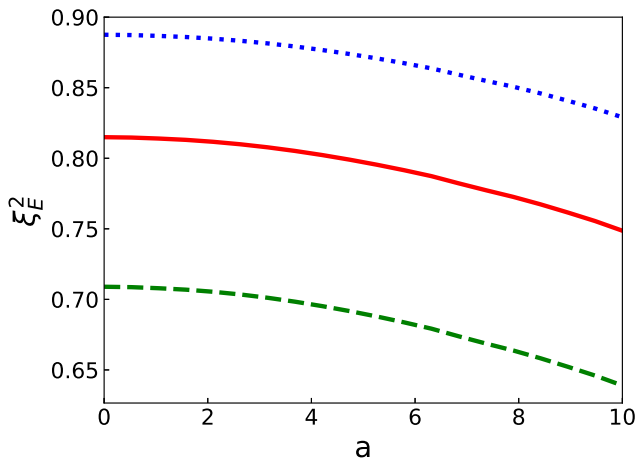} \caption{(Color online) The
spin-squeezing parameter as a function of the acceleration $a$ when the atoms
are accelerated in the thermal field. The different lines refer to different
background temperatures: the dashed green line for $T=3$, the solid red line
for $T=4$, and the dotted blue line for $T=5$. We make the total atom number N
= 100, and the other parameters are the same as in Fig. 1. }%
\label{Fig4}%
\end{figure}

\begin{figure}[ptb]
\centering
\includegraphics[width=1\columnwidth]{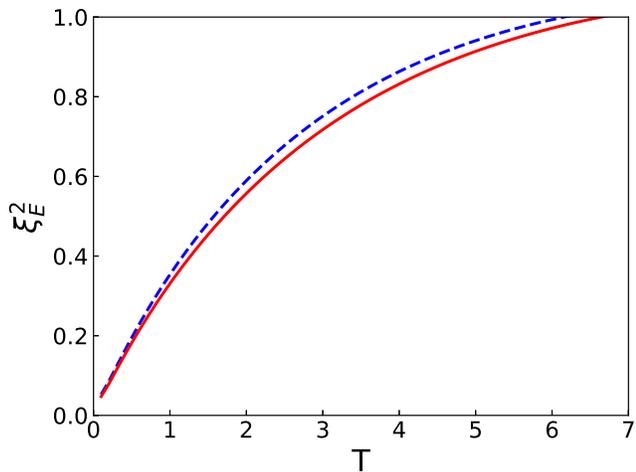} \caption{(Color online) The
spin-squeezing parameter as a function of the background temperature when the
atoms are accelerated in the thermal field. The dashed blue line is drawn at
$a = 2.5$, and the solid red line at $a = 7.5$ We make the total atom number N
= 100, and the other parameters are the same as in Fig. 1. }%
\label{Fig5}%
\end{figure}

\section{Phase Sensitivity}

In this section, we continue to discuss the influence of background
temperature on entanglement with the actual experimental conditions. For this
purpose, we have to study the influence of acceleration on the phase
sensitivity and compare it with the present experiment, since the change of
spin squeezing or entanglement influences the phase sensitivity of the measurement.

Consider the Ramsey interferometer \cite{nr85,ymk86} with the initial input
state $\rho_{i}$, and the output state $\rho_{o}=U^{\dag}\rho_{i}U$ where
$U=\exp\left(  -i\theta J_{y}\right)  $ is the unitary operator for the
evolution and $\theta$ is the phase shift. The phase sensitivity $\Delta
\theta$ can be calculated as%
\begin{equation}
\left(  \Delta\theta\right)  _{P}^{2}=\frac{2\left(  \Delta J_{z}^{2}\right)
_{i}\left(  \Delta J_{x}^{2}\right)  _{i}+V_{xz}}{4\left(  \left\langle
J_{x}^{2}\right\rangle _{i}-\left\langle J_{z}^{2}\right\rangle _{i}\right)
^{2}}, \label{ops}%
\end{equation}
where the subscript $i$ denotes that the average is taken under the input
state and $V_{xz}=\left\langle \left(  J_{x}J_{z}+J_{z}J_{x}\right)
^{2}\right\rangle _{i}+\left\langle J_{z}^{2}J_{x}^{2}+J_{x}^{2}J_{z}%
^{2}\right\rangle _{i}-2\left\langle J_{z}^{2}\right\rangle _{i}\left\langle
J_{x}^{2}\right\rangle _{i}$. This is obtained from the error propagation
formula, $\left(  \Delta\theta\right)  ^{2}=\frac{\left(  \Delta J_{z}%
^{2}\right)  _{o}^{2}}{\left\vert d\left\langle J_{z}^{2}\right\rangle
_{o}/d\theta\right\vert ^{2}}$, by using the relation $U^{\dag}J_{z}%
U=J_{z}\cos\theta-J_{x}\sin\theta$ to link the input state with the output
state and by choosing the optimal phase shift through $\tan^{2}\theta
_{p}=\frac{\left(  \Delta J_{z}^{2}\right)  _{i}}{\left(  \Delta J_{x}%
^{2}\right)  _{i}}$. For the accelerated state in Eq. (\ref{dsm}), a direct
but tedious calculation within the approximation, $m<<j$ and $A_{m}^{2}<<$
$A_{0}^{2}$ gives%
\begin{align}
\left(  \Delta\theta\right)  _{PA}^{2}  &  \simeq\frac{1}{2j\left(
j+1\right)  }+\frac{2\left\langle J_{z}^{2}\right\rangle }{j\left(
j+1\right)  }\nonumber\\
&  +\frac{\sum_{m=-N/2}^{N/2}\sqrt{2}\Delta J_{z}^{2}}{2j\left(  j+1\right)
}, \label{psr}%
\end{align}
which is obtained from Eq. (\ref{ops}) with $\left\langle J_{z}^{2}%
\right\rangle =\sum_{m=-N/2}^{N/2}m^{2}A_{m}^{2}$, $\Delta J_{z}^{2}%
=\sum_{m=-N/2}^{N/2}m^{4}A_{m}^{2}-\left(  \sum_{m=-N/2}^{N/2}m^{2}A_{m}%
^{2}\right)  ^{2}$, $\left\langle J_{x}^{2}\right\rangle =\frac{1}{2}\left[
j\left(  j+1\right)  -\sum_{m=-N/2}^{N/2}m^{2}A_{m}^{2}\right]  $, $\Delta
J_{x}^{2}\simeq\frac{1}{8}\sum_{m=-N/2}^{N/2}A_{m}^{2}j\left(  j+1\right)
\left[  j\left(  j+1\right)  -2-8m^{2}\right]  $, $V_{xz}\simeq\sum
_{m=-N/2}^{N/2}A_{m}^{2}m\left[  \left(  2m+1\right)  ^{2}\alpha_{m}%
^{2}+\left(  2m-1\right)  ^{2}\beta_{m}^{2}\right]  $. The calculation for
$V_{xz}$ keeps to the second order of $m$ since $m$ is small, and $\alpha
_{m}^{2}=\frac{1}{4}\left[  \left(  j-m\right)  \left(  j+m+1\right)  \right]
$, $\beta_{m}^{2}=\frac{1}{4}\left[  \left(  j+m\right)  \left(  j-m+1\right)
\right]  $.

\begin{figure}[ptb]
\centering
\includegraphics[width=1\columnwidth]{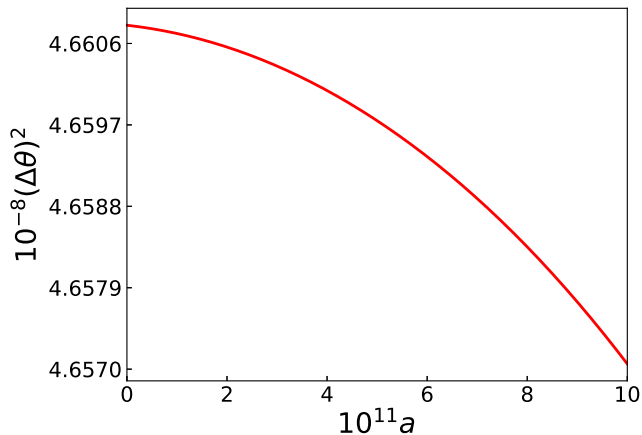} \caption{(Color online) The
phase sensitivity as a function of the acceleration $a$. We take the
parameters according to the experiment made in Ref. \cite{lzy17} with
$\lambda=1$, $\sigma=26(\mu s)$, $\Omega=2\pi\hbar$, and $N=10000$.}%
\label{Fig6}%
\end{figure}

With the formula (\ref{psr}), we numerically calculate the condition for
appearance of the anti-Unruh effect. Only if the background temperature is
less than $10^{-9}$ K, the anti-Unruh effect can be found. This is
advantageous since the temperature of $10^{-9}$ K is just required to form
Bose-Einstein condensates in the recent experiment that generates the TF state
\cite{lzy17}. At the same time, it is also noted that when the acceleration
reaches 10$^{11}$ m/s$^{2}$, the phase sensitivity is at the level of $\left(
\Delta\theta\right)  _{PA}^{2}\sim10^{-8}$, which requires a more precise
experimental conditions than our earlier estimation \cite{pz20} by the
acceleration in the vacuum, due to the influence of the thermal field.
Although it is not easy to obtain so large acceleration for Bose-Einstein
condensates, the phenomena led by the background temperature is significant
for any other related observations, since the anti-Unruh effect gives the
different behaviors from that given by the thermal background effect, in which
the latter cannot lead to the increase of entanglement. This is attractive for
the future experiment with higher sensitivity. Figure 6 presents the possible
condition of background temperature to realize the case of anti-Unruh effect
with the accelerating TF state. When the background temperature is higher than
$10^{-9}$ K, the anti-Unruh effect would disappear and the Bose-Einstein
condensates would be broken. Similarly, the acceleration can not be larger
than 10$^{11}$ m/s$^{2}$ which corresponds to the Unruh temperature of
$10^{-9}$ K. Of course, it is expected that the observable window for the
Unruh or anti-Unruh effect can occur at smaller acceleration for the future
experimental technologies.

\section{Conclusion}

In this paper, we investigate the influence of the background temperature on
the existence of anti-Unruh effect. At first, we show that for a massless
scalar field the anti-Unruh effect can appear for accelerated detectors in the
thermal background, which is proved at the earlier time that the anti-Unruh
effect is absent for accelerated detectors in the Minkowski vacuum of a
massless scalar field. It is noted that the breakdown of time transitional
invariance of thermal Green function is not crucial for the appearance of
anti-Unruh effect. We have also calculated the change of entanglement for
many-body quantum states accelerating in thermal field. It is found that the
anti-Unruh effect still exists, but it will change when the background
temperature increases while all other parameters are not changed. In
particular, when the background temperature increases to a specific value,
entanglement would decrease to zero, which wipes off any possibility to
observe the influence of acceleration on entanglement irrespective of the case
for the Unruh effect or the anti-Unruh effect. Finally, we have used the
experimental parameters to estimate the possibility for the existence of the
anti-Unruh effect. It is interesting to find that the case for anti-Unruh
effect can appear for such accelerated states only if the background
temperature is less than $10^{-9}$K, which is favorable for the possible
future experiment since this effect is distinctive and different from that
coupled to a thermal environment directly by inertial detectors.

\section{Acknowledgement}

This work is supported from Grant No. 11654001 of the National Natural Science
Foundation of China (NSFC).

\end{document}